\documentclass[aps,prl,superscriptaddress,twocolumn,showpacs]{revtex4}%,twocolumn

\usepackage{epsfig}
\usepackage{amsmath}
\usepackage{times}
\usepackage{graphicx}
\usepackage{amssymb}
\usepackage{multirow}

\begin{document}

\title{Predicting catastrophes in nonlinear dynamical systems by compressive sensing}

\author{Wen-Xu Wang}
\affiliation{School of Electrical, Computer and Energy
Engineering, Arizona State University, Tempe, AZ 85287}

\author{Rui Yang}
\affiliation{School of Electrical, Computer and Energy
Engineering, Arizona State University, Tempe, AZ 85287}

\author{Ying-Cheng Lai}
\affiliation{School of Electrical, Computer and Energy
Engineering, Arizona State University, Tempe, AZ 85287}
\affiliation{Institute for Complex Systems and Mathematical
Biology, King's College, University of Aberdeen, Aberdeen AB24
3UE, UK}

\author{Vassilios Kovanis}
\affiliation{Electro-Optics Components Branch,Sensors Directorate,
2241 Avionics Circle,Wright Patterson AFB, OH 45433}

\author{Celso Grebogi}
\affiliation{Institute for Complex Systems and Mathematical
Biology, King's College, University of Aberdeen, Aberdeen AB24
3UE, UK}

\date{\today}

\begin{abstract}

An extremely challenging problem of significant interest is to
predict catastrophes in advance of their occurrences. We present a
general approach to predicting catastrophes in nonlinear dynamical
systems under the assumption that the system equations are
completely unknown and only time series reflecting the evolution
of the dynamical variables of the system are available. Our idea
is to expand the vector field or map of the underlying system into
a suitable function series and then to use the compressive-sensing
technique to accurately estimate the various terms in the
expansion. Examples using paradigmatic chaotic systems are
provided to demonstrate our idea.

\end{abstract}

 \maketitle

It has been recognized that nonlinear dynamics are ubiquitous in
many natural and engineering systems. A nonlinear system, in its
parameter space, can often exhibit catastrophic bifurcations that
ruin the desirable or ``normal'' state of operation. Consider, for
example, the phenomenon of crisis \cite{GOY:1983} where, as a
system parameter is changed, a chaotic attractor collides with its
own basin boundary and is suddenly destroyed. After the crisis,
the state of the system is completely different from that on the
attractor before the crisis. Suppose that, for a nonlinear
dynamical system, the state before the crisis is normal and
desirable, and the state after the crisis is undesirable or
destructive. The crisis can thus be regarded as a {\em
catastrophe} that one strives to avoid at all cost. Catastrophic
events, of course, can occur in different forms in all kinds of
natural and man-made systems. A question of paramount importance
is how to predict catastrophes in advance of their possible
occurrences. This is especially challenging when the equations of
the underlying dynamical system are unknown and one must then rely
on measured time series or data to predict any potential
catastrophe.

In this paper, we articulate a strategy to address the problem of
predicting catastrophes in nonlinear dynamical systems. We assume
that an accurate model of the system is not available, i.e., the
system equations are unknown, but the time evolutions of the key
variables of the system can be accessed through monitoring or
measurements. Our method consists of three steps: (i) predicting
the dynamical system based on time series, (ii) identifying the
parameters of the system, and (iii) performing bifurcation
analysis using the predicted system equations to locate potential
catastrophic events in the parameter space so as to determine the
likelihood of system's drifting into a catastrophe regime. In
particular, if the system operates at a parameter setting close to
such a critical bifurcation, catastrophe is imminent as a small
parameter change or a random perturbation can push the system
beyond the bifurcation point. To be concrete, in this paper we
regard crises as catastrophes. Once a complete set of system
equations has been predicted and the parameters have been
identified, one needs to examine the available parameter space. In
general, to explore the multi-parameter space of a dynamical
system can be extremely challenging, which can often lead to the
discovery of new phenomena in dynamics. An early example in this
area of research is the work by Stewart et al. \cite{SUGY:1995},
which investigated the phenomenon of double crises in
two-parameter dynamical systems. More recent efforts include the
investigation of hierarchical structures in the parameter space
\cite{Gallas:2008}. The present focus of our work, however, is on
predicting the dynamical systems based on compressive sensing.

The problem of predicting dynamical systems based on time series
has been outstanding in nonlinear dynamics because, despite
previous efforts \cite{KS:1997} in using the standard
delay-coordinate embedding method \cite{Takens:1981} to decode the
topological properties of the underlying system, how to accurately
infer the underlying {\em nonlinear system equations} remains
largely an unsolved problem. In principle, a nonlinear system can
be approximated by a large collection of linear equations in
different regions of the phase space, which can indeed be achieved
by reconstructing the Jacobian matrices on a proper grid that
covers the phase-space region of interest
\cite{FS:1987,Sauer:2004}. However, the accuracy and robustness of
the procedure are challenging issues, including the difficulty
with the required computations. In order to be able to predict
potential catastrophes, local reconstruction of a large set of
linearized dynamics is not sufficient but rather, an accurate
prediction of the underlying nonlinear equations themselves is
needed.

Our framework to fully reconstruct dynamical systems using time
series alone is based on the assumption that the dynamics of many
natural and man-made systems are determined by functions that can
be approximated by series expansions in a suitable base. The major
task is then to estimate the coefficients in the series
representation. In general, the number of coefficients to be
estimated can be large but many of them are zero (the sparsity
condition). According to the conventional wisdom this would be a
difficult problem as a large amount of data is required and the
computations involved can be extremely demanding. However, the
recent paradigm of compressive sensing developed by Cand\`{e}s et
al. \cite{CRT:2006,C:2006,D:2006,B:2007,C:2008} provides a viable
solution to the problem, where the key idea is to reconstruct a
sparse signal from small amount of observations
\cite{CRT:2006,C:2006,D:2006,B:2007,C:2008}, as measured by linear
projections of the original signal on a few predetermined vectors.
Since the requirements for the observations can be considerably
relaxed as compared with those associated with conventional signal
reconstruction schemes, compressive sensing has received much
recent attention and it is becoming a powerful technique to obtain
high-fidelity signal for applications where sufficient
observations are not available. Here, we shall articulate a
general methodology to cast the problems of dynamical-system
prediction into the framework of compressive sensing and we
demonstrate the power of our method by carrying out bifurcation
analyses on the predicted dynamical systems to locate potential
catastrophes using exemplary chaotic systems.

Generally, the problem of compressive sensing can be described as
the reconstruction of a sparse vector ${\bf a}\in R^v$ from linear
measurements ${\bf X}$ about ${\bf a}$ in the form: ${\bf X}={\bf
G} \cdot {\bf a}$, where ${\bf X}\in R^w$, ${\bf G}$ is a $w\times
v$ matrix and most components of ${\bf a}$ are zero. The
compressive sensing theory ensures that the number of components
of the unknown signal can be much larger than the number of
required measurements for reconstruction, i.e., $v\gg w$. Accurate
reconstruction can be achieved by solving the following convex
optimization problem \cite{CRT:2006}: $\min \|{\bf a}\|_1 \quad
\mbox{subject \ to} \quad {\bf X}={\bf G} \cdot {\bf a}$, where
$\|{\bf a}\|_1=\sum_{i=1}^{v}|{\bf a}_i|$ is the $L_1$ norm of
${\bf a}$. Solutions to the convex optimization problem have been
worked out recently
\cite{CRT:2006,C:2006,D:2006,B:2007,C:2008,CR:2005}.

We first show that the inverse problem of predicting dynamical
systems can be cast in the framework of compressive sensing so
that optimal solutions can be obtained even when the number of
base coefficients to be estimated is large and/or the amount of
available data is small. In the following, we present a typical
example to illustrate our method. Assume that the dynamical system
can generally be written as $\dot{{\bf x}} = {\bf F}({\bf x})$,
where ${\bf x} \in R^m$ represents the set of externally
accessible dynamical variables and ${\bf F}$ is a smooth vector
function in $R^m$. The $j$th component of ${\bf F}({\bf x})$ can
be represented as a power series:
\begin{eqnarray} \label{eq:2_1}
[{\bf F}({\bf x})]_j = \sum_{l_1=0}^{n}\sum_{l_2=0}^{n}
\cdots\sum_{l_m=0}^{n} (a_j)_{l_1,\cdots,l_m} \cdot x_1^{l_1}
x_2^{l_2} \cdots x_m^{l_m},
\end{eqnarray}
where $x_k$ ($k=1,\cdots, m$) is the $k$th component of the
dynamical variable, and the scalar coefficient of each product
term $(a_j)_{l_1,\cdots,l_m}\in R$ is to be determined from
measurements. Note that the terms in Eq.~(\ref{eq:2_1}) are all
possible products of different components with different powers,
and there are $(1+n)^m$ terms in total.

To better explain our method, without loss of generality, we focus
on one dynamical variable of the system. (Procedures for other
variables are similar.) For example, to construct the measurement
vector ${\bf X}$ and the matrix ${\bf G}$ for the case of $m = 3$
(dynamical variables $x$, $y$, and $z$) and $n = 3$, we have the
following explicit dynamical equation for the first dynamical
variable: $[{\bf F}({\bf x})]_1 \equiv (a_1)_{0,0,0}x^0y^0z^0 +
(a_1)_{1,0,0}x^1y^0z^0 + \cdots + (a_1)_{3,3,3}x^3y^3z^3$. We can
denote the coefficients of $[{\bf F}({\bf x})]_1$ by ${\bf a}_1 =
[(a_1)_{0,0,0},(a_1)_{1,0,0},\cdots ,(a_1)_{3,3,3}]^T$. Assuming
that measurements of ${\bf x}(t)$ at a set of time
$t_1,t_2,\ldots,t_w$ are available, we denote ${\bf g}(t) = \big[
x(t)^0y(t)^0z(t)^0, x(t)^0y(t)^0z(t)^1, \cdots, x(t)^3y(t)^3z(t)^3
\big]$, such that $[{\bf F}({\bf x}(t))]_1 = {\bf g}(t) \cdot {\bf
a}_1$. From the expression of $[{\bf F}({\bf x})]_1$, we can
choose the measurement vector as $\mathbf{X}
=\left[\dot{x}(t_1),\dot{x}(t_2), \cdots,\dot{x}(t_w)\right]^T$,
which can be calculated from time series. Finally, we obtain the
following equation in the form $\mathbf{X} = \mathbf{G} \cdot
\mathbf{a}_1$:
\begin{eqnarray}\label{eq:YeqsAX}
 \left ( \begin{array}{cc}
  \dot{x}(t_1)\\
  \dot{x}(t_2)\\
  \vdots \\
  \dot{x}(t_w)\\
\end{array}\right )=
\left(
\begin{array}{c}
  {\bf g}(t_1) \\
  {\bf g}(t_2) \\
  \vdots  \\
  {\bf g}(t_w) \\
\end{array}
\right ) \left(
\begin{array}{c}
   {\bf a}_1
\end{array}
\right).
\end{eqnarray}
To ensure the restricted isometry property \cite{CRT:2006}, we
normalize ${\bf G}$ by dividing elements in each column by the
$L_2$ norm of that column: $({\bf G'})_{ij} = ({\bf
G})_{ij}/L_2(j)$ with $L_2(j) =\sqrt{ \sum_{i=1}^{M} [({\bf
G})_{ij}]^2}$, so that $\mathbf{X} = \mathbf{G'} \cdot
\mathbf{a}_1'$. After the normalization, ${\bf a_1'}={\bf
a_1}\cdot L_2$ can be determined via some standard
compressive-sensing algorithm \cite{CR:2005}. As a result, the
coefficients ${\bf a_1}$ are given by ${\bf a_1'}/L_2$. To
determine the set of power-series coefficients corresponding to a
different dynamical variable, say $y$, we simply replace the
measurement vector by ${\bf X} =\left[\dot{y}(t_1),\dot{y}(t_2),
\cdots,\dot{y}(t_w)\right]^T$ and use the same matrix ${\bf G}$.
This way all coefficients can be estimated.

\begin{figure}
\begin{center}
\epsfig{file=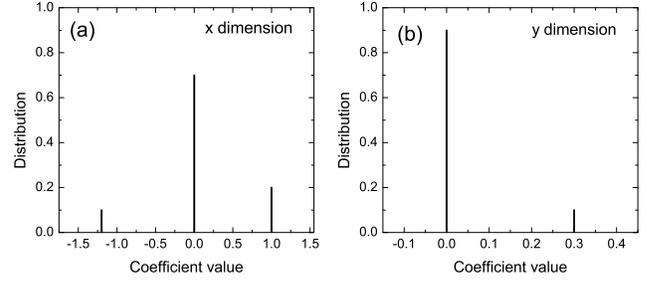,width=\linewidth} \caption{For the
H\'{e}non map, in (a) $x$ dimension and (b) $y$ dimension,
distributions of the predicted values of ten power-series
coefficients up to order 3: constant, $y$, $y^2$, $y^3$, $x$,
$xy$, $xy^2$, $x^2$, $x^2y$ and $x^3$.} \label{fig:distribution}
\end{center}
\end{figure}

\begin{figure}
\begin{center}
\epsfig{file=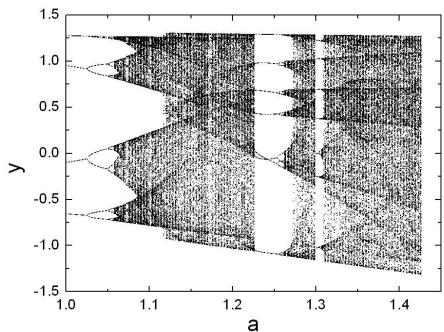,width=0.7\linewidth} \caption{Bifurcation
diagrams of the predicted H\'enon map. The predicted map equations
are: $x_{n+1} = 0.999999996105743+1.000000008610316 \times y_n - a
x_n^2$ and $y_{n+1}=0.29999999837 \times x_n$. The number $n_m$ of
measurements used for prediction is 8 and the total number
$n_{nz}+n_z$ of terms to be predicted is 16.} \label{fig:henon}
\end{center}
\end{figure}

\begin{figure}
\begin{center}
\epsfig{file=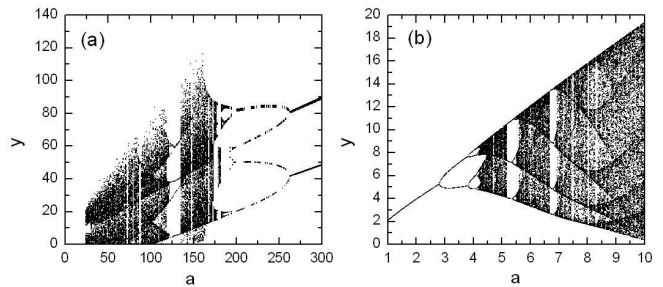,width=\linewidth} \caption{Bifurcation
diagram of (a) the predicted Lorenz system given by
$\dot{x}=10.000548307148881\times y - 10.001897147696283\times x$,
$\dot{y} = x(a- 1.000933186801829\times z)-1.000092963203845\times
y$, $\dot{z}=0.999893761636553\times xy - 2.666477325955504\times
z$ and of (b) the predicted R\"ossler system given by $\dot{x} =
-0.999959701293536\times y - 0.999978902248041\times z$, $\dot{y}
= 1.000004981649221\times x + 0.200005996113158\times y$, $\dot{z}
= 0.199997011085648 + 0.999999156496251\times z( x - a)$. In both
cases, $n_m = 18$ and $n_{nz} + n_z = 35$. }
\label{fig:Lorenz_Rossler}
\end{center}
\end{figure}

We now present a number of physically relevant examples to
illustrate our strategy. The first example is the H\'{e}non map
\cite{Henon:1976}, a classical model that has been used to address
many fundamental issues in chaotic dynamics. The prediction of map
equations resembles that of a vector field. The map is given by:
$(x_{n+1},y_{n+1}) = (1 - a x_n^2 + y_n, bx_n)$, where $a$ and $b$
are parameters. For $b = 0.3$, the map exhibits periodic and
chaotic attractors for $a < a_c \approx 1.42625$, where $a_c$ is
the critical parameter value for a boundary crisis
\cite{GOY:1983}, above which almost all trajectories diverge. The
crisis can then be regarded as a catastrophe in the system
evolution. Assuming, e.g., that the ``normal'' operation of the
system corresponds to a chaotic attractor, we choose $a = 1.4$.
Now suppose that the system operates at this parameter value and
the system equations are completely unknown but the time series
$(\{x\}_n,\{y\}_n)$ can be obtained in real time. The goal is to
assess, based on the time series only, how ``close'' the system is
to a potential catastrophe. (If measurements of only one dynamical
variable can be obtained, one has to resort to the
delay-coordinate embedding method \cite{Takens:1981}.) For
illustrative purpose, we assume power-series expansions up to
order 3 in the map equations. Figure \ref{fig:distribution} shows
the distributions of the estimated power-series coefficients,
where we observe extremely narrow peaks about zero, indicating
that a large number of the coefficients are effectively zero,
which correspond to nonexistent terms in the map equations.
Coefficients that are not included in the zero peak correspond
then to existent terms and they determine the predicted map
equations. Figure \ref{fig:henon} shows the bifurcation diagram
from the predicted H\'enon map, which is consistent with the
original diagram impressively well. In particular, the predicted
system gives the value of the critical bifurcation point to within
$10^{-3}$, where a boundary crisis occur. Note that, to predict
correctly the map equations, the number of required data is
extremely low, not seen before in any method of dynamical-system
reconstruction. Similar results have been obtained for the chaotic
Lorenz \cite{Lorenz:1963} and R\"{o}ssler \cite{Rossler:1976}
oscillators, as shown by the predicted bifurcation diagrams in
Figs. \ref{fig:Lorenz_Rossler}(a) and \ref{fig:Lorenz_Rossler}(b),
respectively. These agree with the original bifurcation diagrams
extremely well, so that all possible critical bifurcation points
can be predicted accurately based on time series only.

\begin{figure}
\begin{center}
\epsfig{file=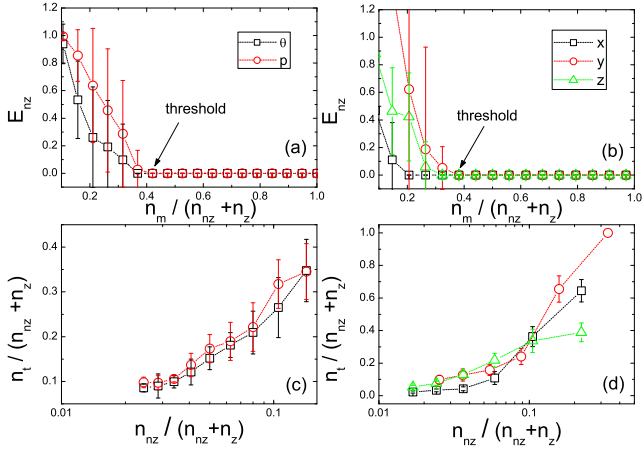,width=\linewidth} \caption{Prediction
errors $E_{nz}$ in dynamical equations as a function of the ratio
of the number $n_m$ of measurements to the total number
$n_{nz}+n_z$ of terms to be predicted for (a) the standard map and
(b) the Lorenz system. The ratio of the threshold $n_t$ to
$n_{nz}+n_z$ for different equations as a function of the ratio
$n_{nz}/(n_{nz}+n_z)$ for (c) the standard map and (d) the Lorenz
system. In (a) and (b), $n_{nz}+n_z$ is 20 and 35, respectively.
The error bars represent the standard deviations obtained from 30
independent realizations. In (c) and (d), $n_{nz}+n_z$ can be
adjusted by the order of power series. In (c),the data points
ranges from order 3 to order 11, and in (d) from order 2 to order
7. We find that $E_{nz}$ and $E_z$ exhibit the same threshold.
}\label{fig:measurement}
\end{center}
\end{figure}

To quantify the performance of our method with respect to the
amount of required data, we investigate the prediction errors
which are defined separately for nonzero (existing) and zero terms
in the dynamical equations. The relative error of a nonzero term
is defined as the ratio to the true value of the absolute
difference between the predicted and true values. The average over
the errors of all terms in a component is the prediction error
$E_{nz}$ of nonzero terms for the component. In contrast, the
absolute error $E_z$ is used for zero terms. Figures
\ref{fig:measurement}(a) and \ref{fig:measurement}(b) show
$E_{nz}$ as a function of the ratio of the number $n_m$ of
measurements to the total number $n_{nz} + n_z$ of terms to be
predicted, for the standard map \cite{Stardard:1979} and the
Lorenz system, respectively. Note that, for the standard map, it
is necessary to explore alternative bases of expansion so that the
sparsity condition can be satisfied. Our strategy is that,
assuming a rough idea about the basic physics of the underlying
dynamical system is available, we can choose a base that is
compatible with the knowledge. In the case of the standard map, we
thus choose the base which includes the trigonometric functions.
We obtain that, when the number $n_m$ of measurements exceeds a
threshold $n_t$, $E_{nz}$ becomes effectively zero. Without loss
of generality, we define $n_t$ by using the small threshold value
$E_{nz} = 10^{-3}$ so that $n_t$ is the minimum number of required
measurements for an accurate prediction. In Figs.
\ref{fig:measurement}(a) and \ref{fig:measurement}(b), we observe
that $n_t$ is much less than $n_{nz} + n_z$ if $n_{nz}$, the
number of nonzero terms, is small. The performance of our method
can thus be quantified by the threshold with respect to the
numbers of measurements and terms to be predicted. As shown in
Figs. \ref{fig:measurement}(c) and \ref{fig:measurement}(d) for
the standard map and the Lorenz system, respectively, as the
nonzero terms become sparser among all terms to be predicted
(characterized by a decrease in $n_{nz}/(n_{nz}+n_z)$ when
$n_{nz}+n_z$ is increased), the ratio of the threshold $n_t$ to
the total number of terms $n_{nz}+n_z$ becomes smaller. These
results demonstrate the advantage of our compressive-sensing based
method to predict dynamical systems, i.e., high accuracy and
extremely low required measurements. In general, to predict the
nonlinear dynamical system as accurately as possible, many
reasonable terms should be assumed in the expansions, insofar as
the percentage of nonzero terms is small so that the sparsity
condition of compressive sensing is satisfied.

\begin{figure}
\begin{center}
\epsfig{file=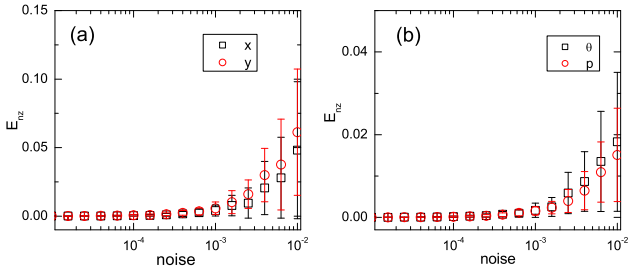,width=\linewidth} \caption{Prediction
errors $E_{nz}$ as a function of noise amplitude for (a) the
H\'enon map and (b) the standard map. Uniform noise is added to
the time series. The values of $n_m$ and $n_{nz} + n_z$ for (a)
are 8 and 16, respectively, and for (b) are 10 and 20,
respectively. The prediction errors in the zero terms show similar
behaviors.} \label{fig:noise}
\end{center}
\end{figure}

In addition, we examine the resistance of the method to
measurement errors by inserting noise into time series. The
prediction errors as a function of noise amplitude are shown in
Figs. \ref{fig:noise}(a) and \ref{fig:noise}(b) for the H\'enon
map and the standard map, respectively. The results demonstrate
that our method is robust against noise, due to the optimization
nature of the compressive-sensing method.

There are also situations where the system is high-dimensional or
stochastic, for which the current method may not work. A possible
solution is to employ the Bayesian inference to determine the
system equations. In general the computational challenge
associated with the approach can be formidable, but the
power-series or more general expansion based compressive-sensing
method developed in this paper may present an effective strategy
to overcome the difficulty.

In summary, we have articulated a general approach to predicting
catastrophes in nonlinear dynamical systems. Our idea is to
approximate the equations of the underlying system by series
expansion and then to formulate the problem of estimating the
various terms in the expansions using compressive sensing. The
merit of our approach is then that, due to the nature of the
compressive-sensing method, a large number of terms can be
accurately estimated even with short available time series,
enabling potential implementation in real times. We have presented
a number of examples from chaotic dynamics to demonstrate the
effectiveness of our method. Predicting catastrophe is a problem
of uttermost importance in science and engineering and of
extremely broad interest as well, and we hope our work will
stimulate further efforts in this challenging area.

This work was supported by AFOSR under Grants No. FA9550-10-1-0083
and FA9550-09-1-0260.
%CG was supported by BBSRC
%under Grants No. BB-F00513X and No. BB-G010722.

\end{document}